\documentclass{osa-article}

\journal{oe}

\articletype{Research Article}

\usepackage{lineno}
\usepackage{braket}
\newcommand{\gpe}{g_{\text{pe}}}
\newcommand{\gom}{g_{\text{om}}}
\newcommand{\gomPE}{g_{\text{om,PE}}}
\newcommand{\gomMB}{g_{\text{om,MB}}}
\newcommand{\etaom}{\eta_{\text{om}}}
\newcommand{\gammaom}{\gamma_{\text{om}}}
\newcommand{\etaext}{\eta_{\text{ext}}}
\newcommand{\etape}{\eta_{\text{pe}}}

\begin{document}

\title{Design of an ultra-low mode volume piezo-optomechanical quantum transducer} 


\author{Piero Chiappina,\authormark{1, 2} Jash Banker,\authormark{1, 2} Srujan Meesala \authormark{1, 2}, David Lake \authormark{1, 2}, Steven Wood \authormark{1, 2}, and Oskar Painter\authormark{1,2,3, *}}

\address{\authormark{1} Thomas J. Watson Sr. Laboratory of Applied Physics, California Institute of Technology, Pasadena, CA, 91125, USA\\
\authormark{2} Kavli Nanoscience Institute and Institute for Quantum Information and Matter, California Institute of Technology, Pasadena, CA, 91125, USA\\
\authormark{3}AWS Center for Quantum Computing, Pasadena, CA, 91125, USA}

\email{\authormark{*}opainter@caltech.edu} 



\begin{abstract}


Coherent transduction of quantum states from the microwave to the optical domain can play a key role in quantum networking and distributed quantum computing. We present the design of a piezo-optomechanical device formed in a hybrid lithium niobate on silicon platform, that is suitable for microwave-to-optical quantum transduction. Our design is based on acoustic hybridization of an ultra-low mode volume piezoacoustic cavity with an optomechanical crystal cavity. The strong piezoelectric nature of lithium niobate allows us to mediate transduction via an acoustic mode which only minimally interacts with the lithium niobate, and is predominantly silicon-like, with very low electrical and acoustic loss. We estimate that this transducer can realize an intrinsic conversion efficiency of up to $35$\% with $<0.5$ added noise quanta when resonantly coupled to a superconducting transmon qubit and operated in pulsed mode at 10 kHz repetition rate. The performance improvement gained in such hybrid lithium niobate-silicon transducers make them suitable for heralded entanglement of qubits between superconducting quantum processors connected by optical fiber links.
\end{abstract}

\section{Introduction}
\label{sec:1}
Recent landmark demonstrations have established superconducting quantum circuits as a leading platform for quantum computing \cite{USTCQAdvantage, GoogleSupremacy, YaleBreakeven2023, GoogleQEC}. Superconducting processors encode quantum information in microwave-frequency photons and require operation at milliKelvin temperatures. Due to the high propagation loss and thermal noise of microwave links at room temperature, transmitting quantum information from superconducting processors over long distances remains an outstanding challenge. In contrast, optical photons are naturally suited for low loss, long distance transmission of quantum information \cite{600kmCommunication, 50kmCommunication}. The complementary properties of these two systems have spurred interest in transducers that can coherently convert quantum information between microwave and optical frequencies. Such transducers would enable optically connected networks of remote superconducting quantum processors analogous to classical networks underlying the internet and large-scale supercomputers with optical interconnects. 

A quantum transducer operated as a frequency converter can be specified as a linear device with a certain conversion efficiency, added noise level, and repetition rate for pulsed operation. Current approaches for microwave-to-optical frequency conversion rely on a strong optical pump to mediate the conversion process between single photon-level signals at both frequencies \cite{JILA_Nondestructive, JFink_Transduction, JFink_Entanglement, HTang_Transduction_EO_AlN, Loncar_Transduction, PainterTransduction, ASN_Transduction, ASN_HeraldedPhotons, HTang_Transduction_OM, KartikTransduction_Nature, QPhoX_Transduction, ClelandTransduction, MM_Transduction, Faraon_Transduction, HongTangTransductionReview, PerspectivesOnQuantumTransduction}. Increasing pump power allows for higher conversion efficiency, but due to parasitic effects of optical absorption in various components of the transducer, and the vast difference in energy scales between optical and microwave frequencies, this adds noise to the conversion process. For quantum applications of the transducer, the number of added noise photons per up-converted signal photon should be $\lesssim 1$ \cite{FiguresOfMerit}. This trade-off between efficiency and noise has been a key obstacle to transduction of quantum signals. 

One particularly promising approach to microwave-to-optical transduction is piezo-optomechanics \cite{PainterTransduction, ASN_HeraldedPhotons, ASN_Transduction, QPhoX_Transduction, KartikTransduction_Nature, ClelandTransduction, HTang_Transduction_OM}, in which acoustic phonons are used as intermediate excitations in the conversion process. This is achieved through a highly engineered mechanical mode with simultaneous piezoelectric and optomechanical couplings. 
Recent design and materials advances in these devices have led to a demonstration of optical readout of the quantum state of a superconducting qubit with added noise below one photon \cite{PainterTransduction}. However, the aluminum nitride piezoelectric element contributed significantly to acoustic loss, compromising device performance. 

In this work, we propose a new device design on a hybrid material platform, lithium niobate (LN) on silicon, which addresses the limitations of previous work. Our design features a highly miniaturized piezoelectric element so that its negative impact on device performance is negligible, while at the same time maintaining strong piezoelectric coupling rates. This design approach is enabled by the strong piezoelectric coefficients of LN \cite{GaylordLN}. Combined with the excellent optomechanical properties of silicon, our design achieves state-of-the-art piezoelectric and optomechanical performance. While some recent demonstrations have used a lithium niobate-on-silicon material platform \cite{ASN_HeraldedPhotons,QPhoX_Transduction}, the design techniques outlined in this work were not fully exploited in these devices. We show that by leveraging the strengths of the individual materials, our design approach can yield the performance improvements necessary to employ these transducers for remote entanglement of superconducting quantum processors.


\begin{figure}[htbp]
\centering\includegraphics[width=13cm]{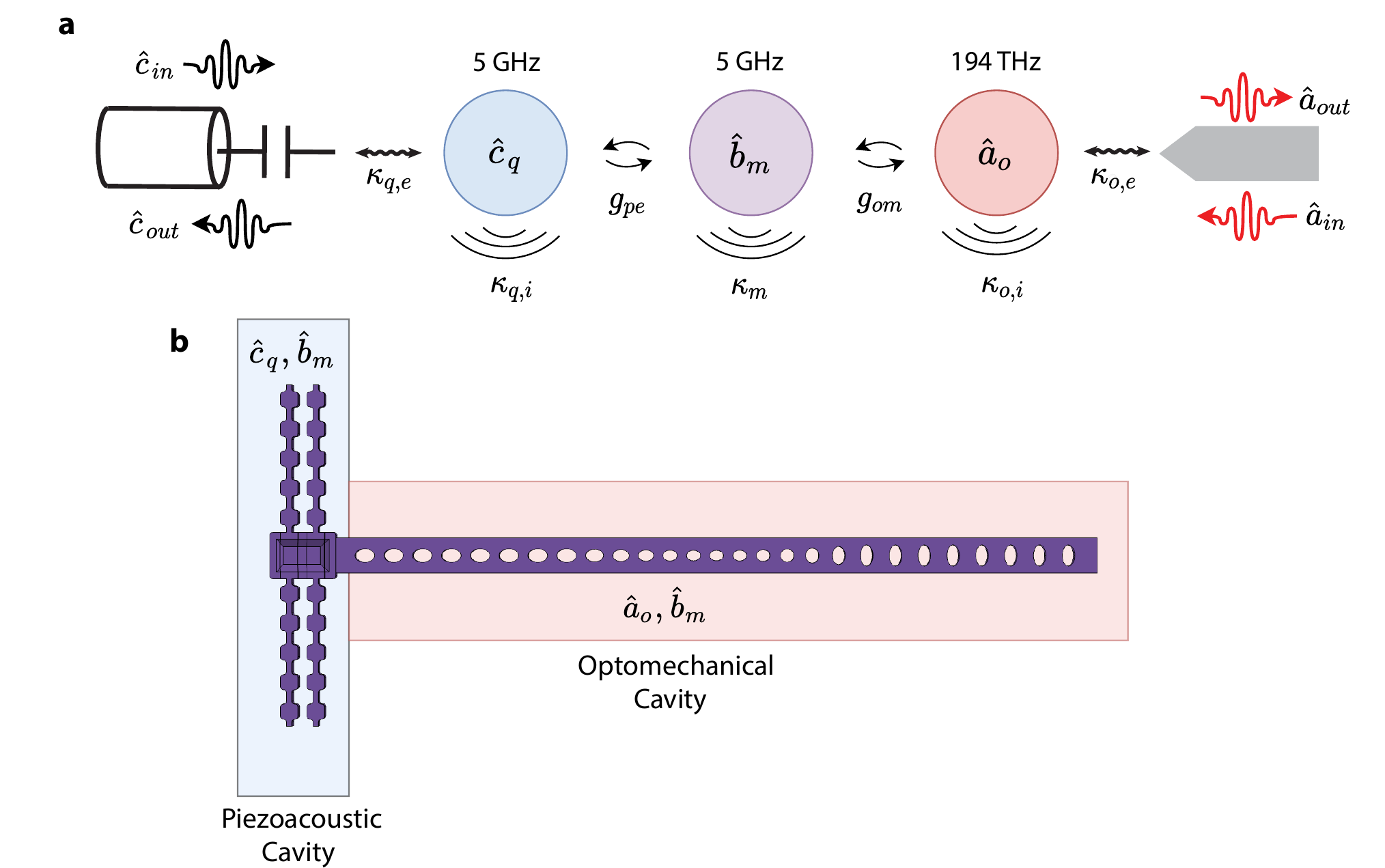}
\caption{a) Mode schematic for piezo-optomechanical transduction. Blue represents mode $\hat{c}_q$ of a microwave circuit, purple represents 'supermode' $\hat{b}_m$ of a mechanical oscillator, and red represents mode $\hat{a}_o$ of an optical cavity. b) Device schematic for the transducer in this work. The device can be split into two regions, one which couples to microwave electric fields and one which couples to optical fields. Both are part of the same mechanical 'supermode' $\hat{b}_m$.}
\label{fig:1}
\end{figure}

Fig.~\ref{fig:1}a illustrates the mode picture of our transduction scheme. An intermediary mode $\hat{b}_m$ of a nanomechanical oscillator simultaneously couples to microwave photons from mode $\hat{c}_q$ of a microwave circuit at rate $\gpe$, and to optical photons from mode $\hat{a}_o$ of an optical cavity at rate $\gom$. Microwave photons are converted to phonons via a resonant piezoelectric interaction, and these phonons are subsequently converted into optical photons via a parametric optomechanical interaction. The microwave photon-phonon conversion is realized by tuning the circuit frequency $\omega_q$ on resonance with the mechanical frequency $\omega_m$, yielding the Hamiltonian $\hat{H}_{\text{pe}}/\hbar =  \gpe(\hat{b}_m^\dagger \hat{c}_q + \hat{b}_m\hat{c}_q^\dagger)$ \cite{YiwenChuAcoustics}. The phonon-optical photon conversion is realized by driving the optical cavity at a frequency $\omega_d$ that is red-detuned by exactly the mechanical frequency, $\omega_{d} - \omega_o = -\omega_m$. The resulting Hamiltonian is $\hat{H}_{\text{om}}/\hbar =  \gom\sqrt{n_o}(\hat{a}_o^\dagger \hat{b}_m + \hat{a}_o\hat{b}_m^\dagger)$ where $n_o$ is the number of intracavity optical photons corresponding to input drive power \cite{CavityOptomechanics}.



We realize the intermediary mechanical mode in the above scheme by connecting an ultra-low mode volume piezoacoustic cavity and an optomechanical crystal (OMC) cavity, shown in Fig.~\ref{fig:1}b. The acoustic modes of these components are strongly hybridized to form a mechanical supermode, whose mechanical displacement overlaps in one region with the field of a microwave circuit, and in another region with the field of an optical cavity. Using physically separate cavities allows us to independently optimize the piezoacoustic and optomechanical components of the transducer. Our design is formed from thin-film LN on the device layer of a silicon-on-insulator (SOI) chip. 
We define the piezoacoustic cavity in LN, which has large piezoelectric coefficients \cite{GaylordLN}. We define the OMC in silicon, since its large photoelastic coefficients and refractive index allow high optomechanical coupling. Well-established nanofabrication processes also allow high optical and mechanical quality factors for silicon OMC's \cite{OMC2009, OMCDesign, GregHighQAcoustics}. For the microwave circuit in this design, we consider a transmon qubit with electrodes routed over the LN region to allow for capacitive coupling to the piezoacoustic cavity. The transmon is patterned in niobium on silicon, a standard material platform for realizing high-coherence qubits \cite{QubitMaterialsReview}. The insulating layer underneath these components is etched away, leaving a suspended silicon membrane as the substrate for our device. 

Our design procedure utilizes finite element simulation in COMSOL Multiphysics \textregistered ~ and numerical optimization of the device geometry. We begin with independently optimizing the piezoacoustic and OMC cavities for high piezoelectric coupling rate $\gpe$ and optomechanical coupling rate $\gom$, respectively. We design for closely matched acoustic modes at 5 GHz in both resonators, and for an optical mode at telecom wavelength (1550 nm). During the design process, it is crucial to maintain a low acoustic mode density such that the transduction schematic in Fig.~\ref{fig:1}a using a single acoustic mode remains valid. Further, since thin film LN has higher microwave dielectric and acoustic loss than silicon \cite{SiLossTangent_LowTemp, LNLossTangent_mK2015, LNLossTangent_mK2023, LNAcousticLossTangent, GregHighQAcoustics}, we aim to minimize the volume of LN in our device. The two independently optimized cavities are then physically connected and the parameters of the resulting hybrid acoustic modes are analyzed.


\section{Piezo Cavity Design}
\label{sec:2}

\begin{figure}[htbp]
\centering\includegraphics[width=13cm]{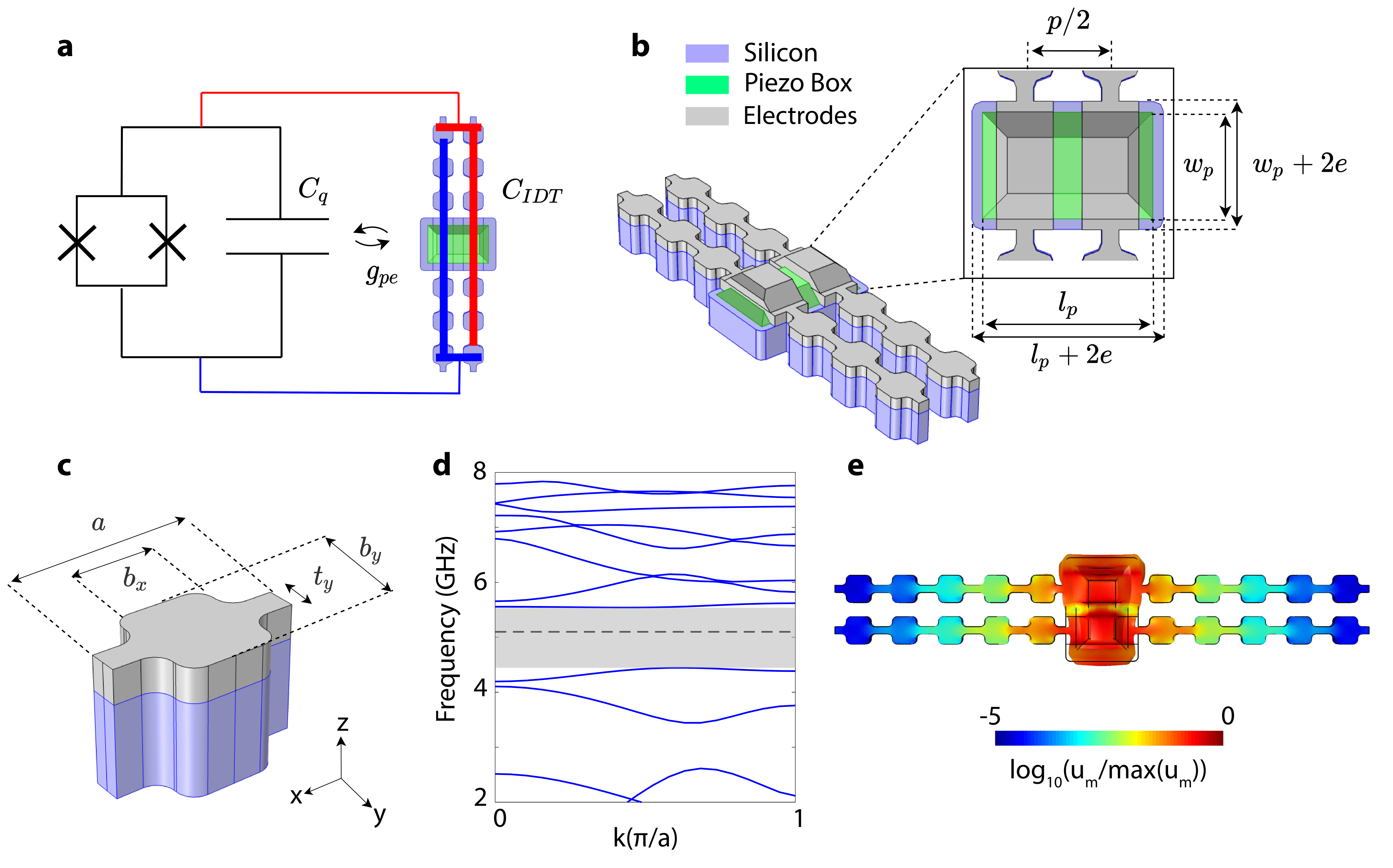}
\caption{a) Schematic illustrating the capacitive routing of the piezoacoustic cavity to a transmon qubit. The qubit here can be replaced with a microwave resonator without loss of generality. b) Piezoacoustic cavity geometry, with relevant dimensions defined in blowout top view. c) Phononic shield unit cell, with relevant dimensions defined. d) Mechanical bandstructure of phononic shield unit cell in c), with $(a, b_x, b_y, t_y) = (445, 225, 265, 70)$~nm. Black dashed line at $5.1$~GHz shows the target frequency of the mode of interest. We observe a complete acoustic bandgap in excess of $1$~GHz around $5.1$~GHz. e) Log scale of mechanical energy density $u_m$ for piezoacoustic cavity mode at $5.1$~GHz, normalized to maximum value. We find $>4$ orders of magnitude suppression for $5$ phononic shield periods.}
\label{fig:2}
\end{figure}

The piezoacoustic cavity consists of a slab of lithium niobate on top of a suspended silicon membrane patterned in the shape of a box. We work with 100~nm thin-film -z-cut LN, on top of a 220~nm-thick suspended silicon device layer. We choose the -z LN orientation so that the dominant LN piezoelectric coefficient, $d_{15} = 69$ pC/N \cite{GaylordLN}, couples to modes with favorable symmetry properties (further discussion in Sec. \ref{sec:3}). 80~nm-thick Nb electrodes run over the top of the slab, and are routed in the form of an interdigital transducer (IDT) which capacitively couples the cavity to a microwave circuit such as a transmon qubit, as seen in Fig.~\ref{fig:2}a. 
 
 The box is surrounded by a periodically patterned phononic shield to mitigate acoustic radiation losses and to clamp the membrane to the surrounding substrate. The clamps are spaced periodically so that the IDT electrodes are routed over the top of each clamp, providing a means for electrical routing which is not acoustically lossy.


The phononic shield uses an alternating block and tether pattern (see relevant dimensions in Fig.~\ref{fig:2}c) consisting of metal electrodes on top of a silicon base. By tuning the parameters $a$, $b_x$, $b_y$, and $t_y$, we achieve a $>1$~GHz acoustic bandgap centered around $5.1$~GHz, shown in Fig.~\ref{fig:2}d. This yields strong confinement of 5~GHz mechanical modes inside the piezo region for sufficient number of shield periods, enabling high mechanical quality factors. By simulating the mechanical energy density across the entire cavity, we find that $5$ shield periods provide $>4$ orders of magnitude suppression of acoustic radiation into the environment, as shown in Fig.~\ref{fig:2}e. 

The dimensions of the piezo box (outlined in Fig.~\ref{fig:2}b) are designed to support a periodic mechanical mode whose periodicity matches that of the IDT fingers. This results in high overlap between the electric field from the IDT and the electric field induced by mechanical motion in the piezo box. This overlap gives a microwave photon-phonon piezoelectric coupling rate which is derived using first order perturbation theory: 

\begin{equation}
\gpe =
 \frac{\omega_m}{4\sqrt{2 U_m U_q}} \int_{\mathrm{LN}} \mathbf{D}_{m} \cdot \mathbf{E}_{q} \hspace{0.1cm} dV.
\label{eq:1}
\end{equation}

\noindent Here the integral is taken over the entire LN slab, $\mathbf{D}_{m}$ is the electric displacement field induced from mechanical motion in the piezo region, and $\mathbf{E}_{q}$ is the single-photon electric field generated by the transmon qubit across the IDT electrodes. The fields are normalized to their respective zero-point energies $\hbar\omega_m/2$, yielding the pre-factor in front of the integral in eq.~(\ref{eq:1}). $U_m$ is the total cavity mechanical energy, and $U_q = \frac{1}{2} (C_q + C_{\mathrm{IDT}}) V_0^2$ is the total IDT electrostatic energy, where $V_{0}$ is the zero-point voltage across the qubit. We note that for fixed electrostatic energy, the zero-point voltage (and thus $\mathbf{E}_{q}$) is dependent on both the qubit capacitance $C_q$ and IDT finger capacitance $C_{\mathrm{IDT}}$, and therefore the coupling rate scales as $(C_q + C_{\mathrm{IDT}})^{-1/2}$. For our calculations in this work, we assume $C_q = 70$~fF which is a typical value for transmon qubit capacitance \cite{TransmonsForEEs}. Replacing the transmon qubit with a high-impedance microwave resonator will allow lower $C_q \sim 2$~fF \cite{NbNResonators, NbNResonators_CPW}, and can therefore further increase this coupling rate. $C_{\mathrm{IDT}}$ is calculated with finite-element electrostatic simulation, and for the wavelength-scale devices considered here is typically $\sim 0.25$~fF, a small contribution compared to transmon $C_q$. 

The small value of $C_{\mathrm{IDT}}$ also minimizes the energy participation of the qubit electric field in the lossy piezo region, given by the ratio $\zeta_q = C_{\mathrm{IDT}}/C_q \sim 4 \times10^{-3}$. 
The contribution of lithium niobate to the qubit loss rate $\kappa_{q, i}$ is then estimated as $\zeta_q \kappa_{q, \mathrm{LN}}$. Using reported dielectric loss tangents in lithium niobate at milliKelvin temperatures, 
$\tan{\delta} = 1.7 \times 10^{-5}$ \cite{LNLossTangent_mK2023, LNLossTangent_mK2015}  corresponding to $\kappa_{q, \mathrm{LN}}/2\pi = 85$~kHz at 5GHz, we estimate the LN contribution to qubit loss to be $\zeta_q \kappa_{q, \mathrm{LN}}/2\pi \sim 300$~Hz. 
This contribution is much smaller than typical loss rates $\kappa_{q, \mathrm{SOI}}/2\pi \sim 50$~kHz reported in transmon qubits fabricated on SOI \cite{AJKQubitsonSOI}.
As a result, the contribution of the piezoacoustic cavity to qubit loss is not a limiting factor, and justifies the on-chip coupling scheme outlined in Fig.~\ref{fig:2}a. 

\begin{figure}[htbp]
\centering\includegraphics[width=13cm]{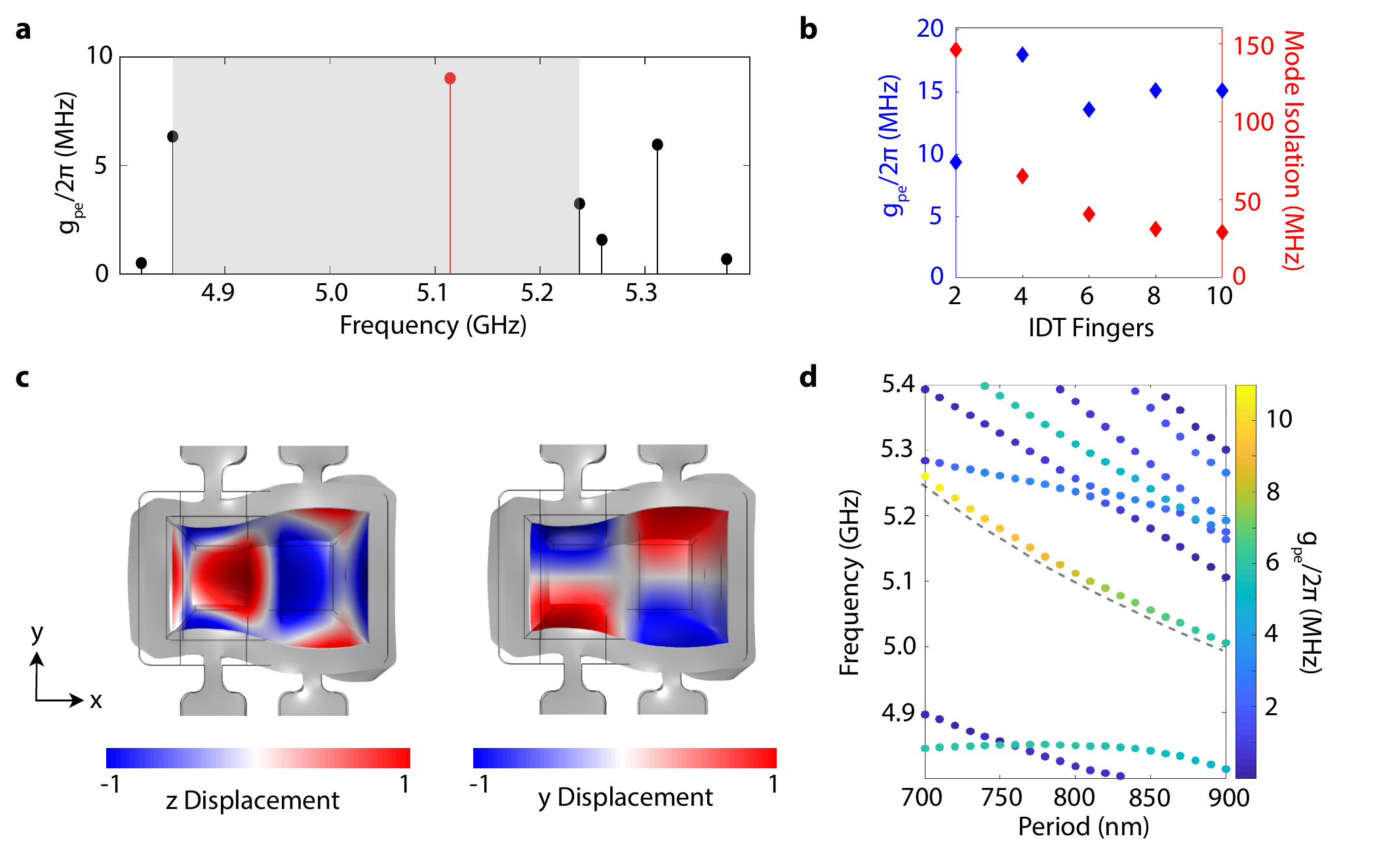}
\caption{a) Mode structure for optimized piezoacoustic cavity design with $(p, w_p, e) = (795, 478, 102)$~nm. Mode of interest (red) achieves $\gpe/2\pi$ of $9$~MHz. Shaded grey region indicates the mode isolation window with the nearest mode $>110$~MHz away. b) $\gpe/2\pi$ and mode isolation for optimized designs with differing number of IDT fingers. We observe $\gpe$ saturating beyond $N = 4$ fingers, and mode isolation decreasing with increasing number of fingers. c) Mechanical mode shape of red mode in a). Right shows the in-plane (breathing) and left shows the out-of-plane (Lamb-wave) components of the optimized mechanical mode. d) Mode structure vs. IDT period, with data points colored according to $\gpe$. Dashed line follows the mode of interest.}
\label{fig:3}
\end{figure}

We note that the IDT capacitance can be increased by adding more IDT fingers, thereby increasing $\gpe$. 
However, as the number of fingers increases, the increased cavity size results in a more crowded mode structure, and it is more difficult to isolate a single mechanical mode without coupling to parasitic modes in the vicinity of the mode of interest. This is of key importance as these parasitic modes may not hybridize well with the OMC cavity and reduce overall transduction efficiency
. This is shown in Fig.~\ref{fig:3}b, where we simulate the mechanical mode structure vs. the number of IDT fingers in the cavity. We see that $\gpe$ of the mode of interest saturates, while the mode isolation is significantly reduced with increasing IDT fingers. 

Reducing the size of the piezo region is also important to reduce mechanical loss, as lithium niobate has high acoustic loss tangents compared to silicon (further discussion in Sec.~\ref{sec:4}). For these reasons, we choose a 2-finger design to minimize these effects. The strong piezoelectric nature of lithium niobate allows for $\gpe$ values high enough for strong microwave photon-phonon coupling, even in the limit of 2 IDT fingers. We emphasize the small dimensions of the piezoacoustic cavity in this design in contrast with previous work on piezo-optomechanical quantum transducers \cite{KartikTransduction_Nature, KartikTransduction, ASN_Transduction, ClelandTransduction, HTang_Transduction_OM}. The benefits of this approach come at the cost of higher sensitivity of the piezoacoustic modes to changes in cavity dimensions. This can have large effects on hybridization with the OMC cavity and the performance of the final transducer device, which relies on resonant matching of acoustic modes in both regions. We show further in Sec.~\ref{sec:4} that the achievable hybridization between piezo and OMC modes with this small piezo volume approach is large enough to protect the design against typical fabrication disorder.

The mechanical mode of interest is periodic with out-of-plane (Lamb-wave) and in-plane breathing components. The Lamb-wave component of the mode induces an electric field in the piezo region with high overlap with the IDT electric field, while the breathing component of the mode enables hybridization with the breathing mode of the optomechanical crystal to be attached in the full device (see Fig.~\ref{fig:3}c). The piezoacoustic mode can be tuned with three key parameters $p$, $w_p$, and $e$. $p$ is the periodicity of the IDT fingers and is used to parameterize the piezo box length, given by $l_p = Np/2$ where $N$ is the number of IDT fingers. $p$ is used to tune the frequency of the mode of interest while maintaining appropriate phase-matching of the mode periodicity with the IDT fingers (shown in Fig.~\ref{fig:3}d). $w_p$ is the piezo box width, which can be increased to increase $\gpe$ via larger mode volumes or decreased to reduce the mode crowding that results from larger box size. Finally, we define a silicon-piezo buffer parameter $e$, which extends the silicon box length/width by an amount $e$ compared to the piezo box. This buffer is needed to protect against silicon/piezo box misalignment in the fabrication process, and acts as an added degree of freedom for tuning frequency, $\gpe$, and mode isolation. We use numerical optimization to tune parameters $(p, w_p, e)$ to arrive at a design with high piezoelectric coupling and mode isolation. We employ a Nelder-Mead simplex optimization similar to that described in Ref.~\cite{OMCDesign}. After optimization, we obtain the mechanical mode structure shown in Fig.~\ref{fig:3}a. We identify a single mechanical mode with $\gpe/2\pi = 9.01$~MHz, which is isolated by $>110$~MHz from other mechanical modes. We will use this single mode to strongly couple to the modes of an optomechanical crystal cavity to create the mechanical supermode of Fig.~\ref{fig:1}a. 

A point of concern with LN is the in-plane crystal orientation with respect to the IDT electrodes. Due to the strong anisotropy of LN, in general the mode structure and $\gpe$ will vary with this orientation. For -z-cut LN and our choice of mode, both $\gpe$ and frequencies vary minimally with in-plane crystal rotations, with $\gpe$ varying by $<5$\% over a full $2\pi$ rotation. 


\section{Optomechanics Design}
\label{sec:3}

\begin{figure}[htbp]
\centering\includegraphics[width=13cm]{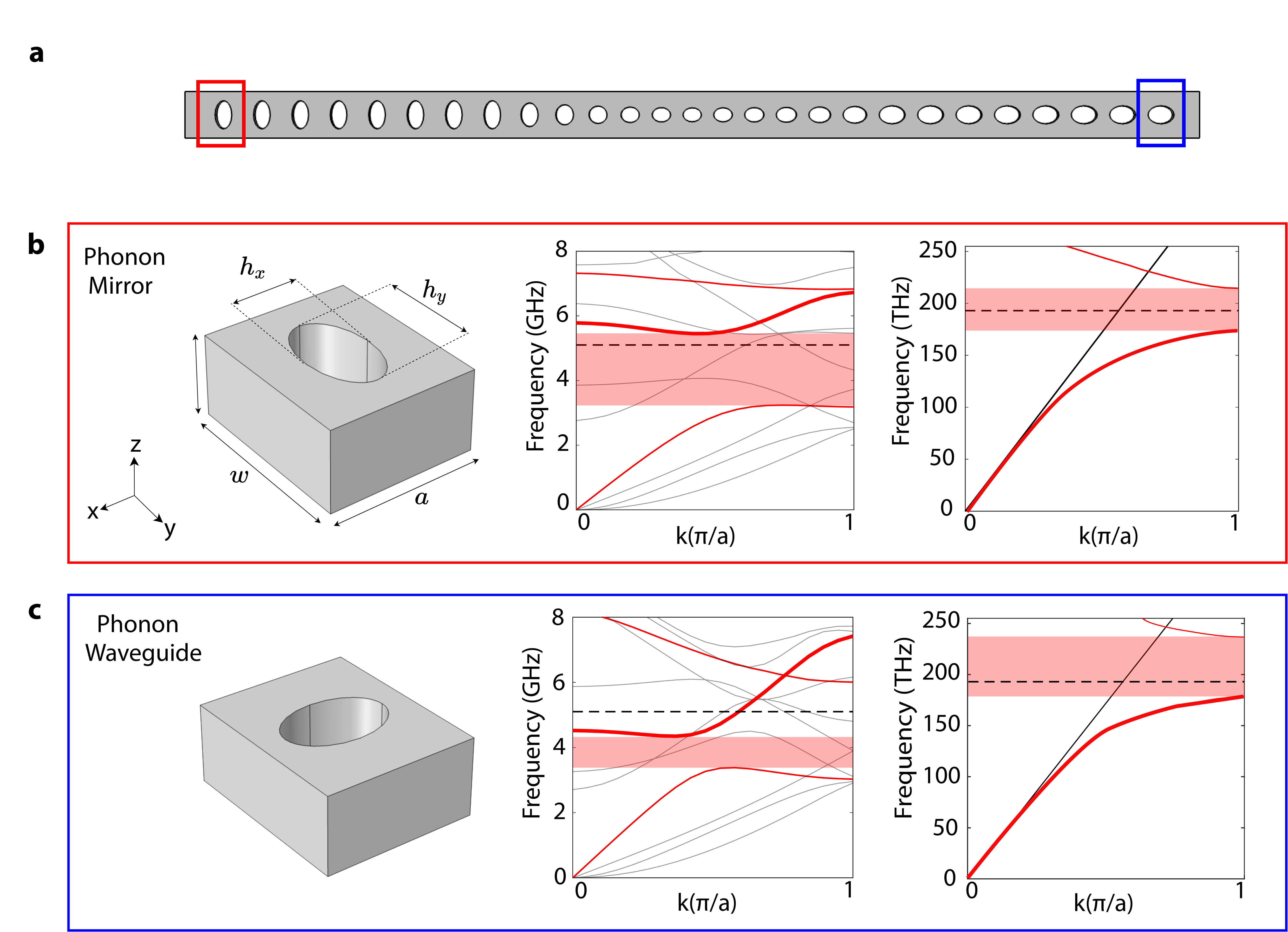}
\caption{a) Full OMC geometry, with phonon mirror and phonon waveguide unit cells highlighted in red and blue, respectively. b) Unit cell geometry, mechanical, and optical bandstructure of phonon mirror region with $(a, w, h_x, h_y) = (436, 529, 181, 334)$~nm. Phonon bands with breathing-mode symmetry are colored in red. Gray bands represent all other symmetries. Mechanical bandgap for breathing modes and optical bandgap are both highlighted in red. Black dashed lines at $5.1$~GHz and $193$~THz show the target frequencies of the 
 phonon/photon modes of interest, respectively.  c) Unit cell geometry, mechanical, and optical bandstructure of phonon waveguide region with $(a, w, h_x, h_y) = (436, 529, 295, 205)$~nm. Mechanical bands color-coded as in a). Breathing mode crosses $5.1$~GHz resulting in waveguide-like behavior at the mechanical frequency. The optical bandgap is maintained.}
\label{fig:4}
\end{figure}

\begin{figure}[htbp]
\centering\includegraphics[width=13cm]{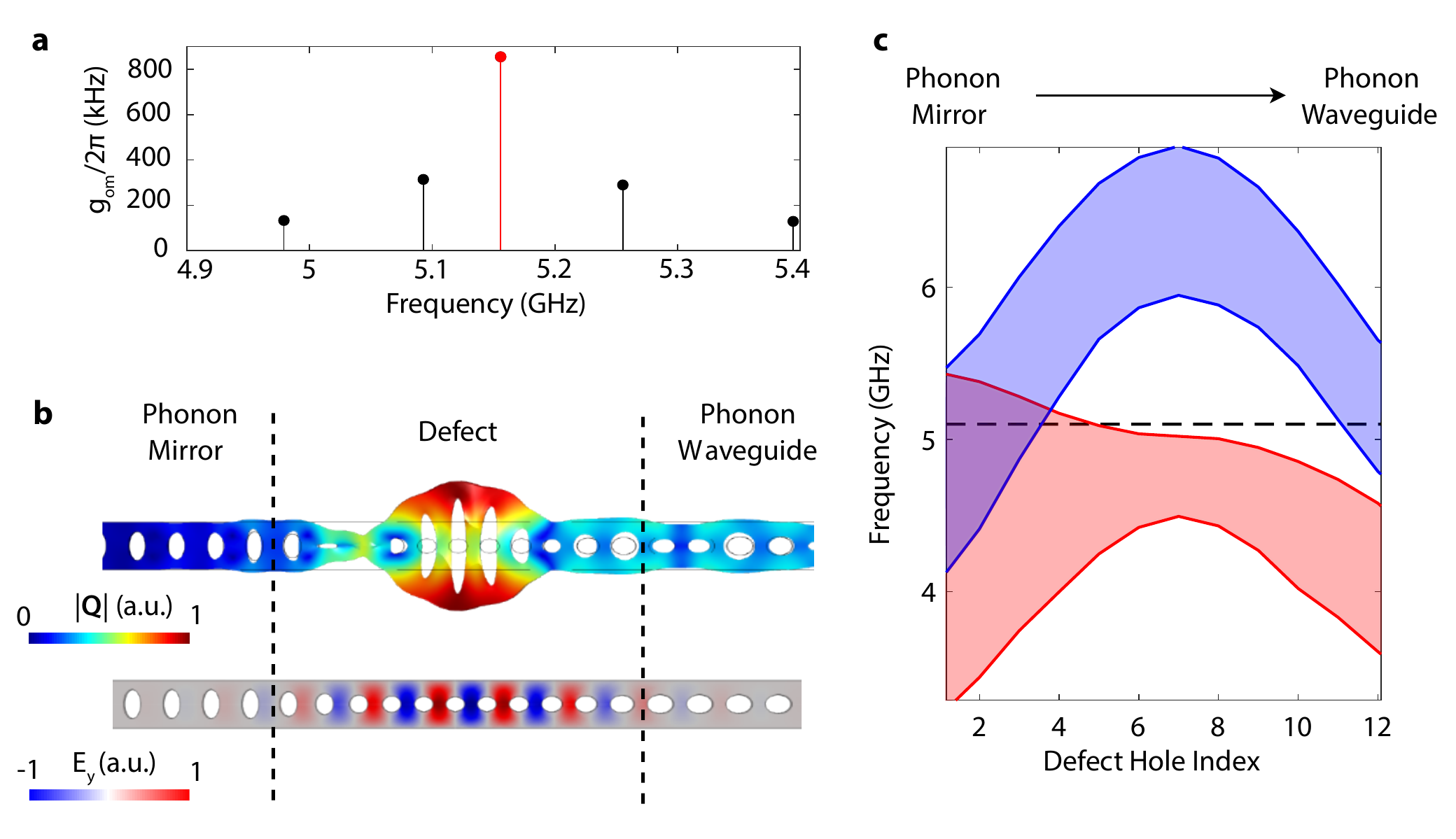}
\caption{a) Resultant mode structure of optimized OMC design. Highest-$\gom$ mode (highlighted in red) gives $854$~kHz optomechanical coupling.  b) Simulated OMC mechanical and optical mode shapes. Top shows mechanical displacement $|\mathbf{Q}|$ and bottom shows y-component of the optical electric field. Phonon mirror, phonon waveguide, and defect region are labeled and outlined. c) Bandgap for modes of different symmetries as a function of hole index along the defect region. Red shaded region represents the bandgap for breathing mode symmetries, and blue shaded region represents the bandgap for Lamb-wave mode symmetries.}
\label{fig:5}
\end{figure}

The optomechanical crystal (OMC) cavity is a periodically patterned silicon nanobeam which is designed in a similar fashion to previous work~\cite{OMCDesign}, with the crucial change of a modified unit cell design on one side of the cavity to enable strong mechanical hybridization with the piezoacoustic cavity. This separates the OMC into three distinct regions: a phonon mirror, defect region, and phonon waveguide (see Fig.~\ref{fig:4} for details). The phonon mirror unit cell (Fig.~\ref{fig:4}b) is designed to have a simultaneous mechanical and optical bandgap for modes of certain symmetry classes. 
In the defect region, the phonon mirror unit cell transitions to a defect cell designed to co-localize a $5.1$~GHz mechanical breathing mode and a $194$~THz ($\lambda_0 = 1550$~nm) optical mode. The phonon waveguide unit cell (Fig.~\ref{fig:4}c) is mechanically transparent to breathing mode phonons at $5.1$~GHz, while maintaining a large bandgap for optical modes. This is achieved by modifying the ellipticity of the phonon mirror unit cell. We see in Fig.~\ref{fig:5}b that the resulting mechanical mode is permitted to leak out into the phonon waveguide region, 
while the optical mode remains highly localized within the defect region. 

The optomechanical coupling rate is calculated from the mechanically-induced optical frequency shift arising due to the photoelastic effect and moving dielectric boundaries~\cite{OMCDesign}, giving $\gom = \gomPE + \gomMB$. The photoelastic contribution is derived from 1st-order perturbation theory as,

\begin{equation}
    \gomPE = \frac{\omega_o \epsilon_0 n^4}{2} \frac{ \int_{\mathrm{Si}} 
    \mathbf{E}^{\dagger} \cdot [\mathbf{p}\mathbf{S}] \cdot \mathbf{E} \hspace{0.1cm} dV}{\int \mathbf{D} \cdot \mathbf{E} \hspace{0.1cm} dV}, \label{eq:2}
\end{equation}

\noindent where $\omega_o$ is the optical cavity frequency, $n$ is the refractive index, $\mathbf{E}$ is the electric field, $\mathbf{p}$ is the photoelastic tensor, and $\mathbf{S}$ is the strain tensor. 
%

 The moving boundaries component is derived similarly as,
 
\begin{equation}
    \gomMB = -\frac{\omega_o}{2} \frac{\oint (\mathbf{Q} \cdot \mathbf{\hat{n}}) (\Delta \epsilon \mathbf{E}_{\parallel}^2 - \Delta \epsilon^{-1} \mathbf{D}_{\perp}^2) dS}{\int \mathbf{D} \cdot \mathbf{E} \hspace{0.1cm} dV}, \label{eq:3}
\end{equation}

\noindent where $\mathbf{Q}$ is the normalized mechanical displacement field, $\mathbf{\hat{n}}$ is the surface normal, $\mathbf{E}_{\parallel}$ is the electric field parallel to the surface, $\mathbf{D}_{\perp}$ is the electric displacement field perpendicular to the surface, $\Delta \epsilon = \epsilon_{\mathrm{Si}} - \epsilon_{\mathrm{Air}}$, and $\Delta \epsilon^{-1} = \epsilon_{\mathrm{Si}}^{-1} - \epsilon_{\mathrm{Air}}^{-1}$. 

One may expect the coupling rates in this design to suffer due to the delocalization of the mechanical mode. However, we find that after a Nelder-Mead simplex optimization of various OMC dimensions similar to \cite{OMCDesign}, the resulting design gives multiple modes with high simulated values of $\gom/2\pi$, with the maximum coupling rate exceeding 850~kHz (Fig.~\ref{fig:5}a). This is comparable to state-of-the-art OMC designs in silicon which achieve $\gom/2\pi$ up to $1.1$~MHz \cite{OMCDesign, SMeenehan, GregHighQAcoustics}. 

The radiation-limited optical quality factor $Q_o$ can be simulated and is found to be in excess of $10^6$, corresponding to an intrinsic optical loss rate $\kappa_{o, i}/2\pi \sim 200$~MHz. However, $Q_o$ is usually practically limited to $\sim 5\times10^5$ ($\kappa_{o, i}/2\pi \sim 400$~MHz) due to optical scattering from surface defects introduced in the fabrication process \cite{OMCDesign}. To ensure this limit is reached, we configure the optimization such that $\gom$ is maximized while maintaining $Q_o$ above $\sim 10^6$, well above the practical $Q_o$ limit. The total optical loss rate is given by $\kappa_o = \kappa_{o, i} + \kappa_{o, e}$, where $\kappa_{o, e}$ is the decay rate associated with input coupling. $\kappa_{o, e}$ is controlled with a coupling waveguide using well-established design techniques \cite{OMC_CouplerDesign}, and is typically designed so that $\kappa_{o, e} = \kappa_{o, i}$. The total optical loss rate is then $\kappa_o \approx 2\kappa_{o, i}/2\pi = 800$~MHz.  

When hybridizing the modes of the piezoacoustic and OMC cavities, we must consider the relative motional symmetry of the two cavity modes. Our OMC cavity design contains only breathing motion, whereas the piezoacoustic cavity design contains both breathing and Lamb-wave components. If the phonon waveguide permits propagation of $5$~GHz Lamb-wave modes, then the OMC defect region will contain both breathing and Lamb-wave motion when hybridized with the piezoacoustic cavity. This will significantly reduce optomechanical coupling as only breathing motion yields high values of $\gom$, whereas Lamb-wave motion produces negligible $\gom$. For this reason, the phonon waveguide dimensions are chosen such that the bandstructure exhibits a bandgap for 5GHz Lamb-wave-like modes, but is transparent to 5GHz breathing modes.

Fig.~\ref{fig:5}c further illustrates this idea by showing the mechanical bandgap of unit cells across the defect region for both breathing and Lamb-wave-like modes, shaded in red and blue respectively. At the phonon waveguide side, the breathing mode bandgap falls below $5$~GHz, permitting the breathing motion of the piezoacoustic mode to couple strongly to the defect region. However, $5$~GHz lies inside the Lamb-wave bandgap, so that Lamb-wave motion from the piezoacoustic cavity decays in the phonon waveguide and does not interact with the OMC cavity defect region. The phonon mirror exhibits a bandgap for both breathing and Lamb wave symmetries, so that acoustic radiation of the full transducer mode to the environment is effectively suppressed.

\section{Full Device Design}
\label{sec:4}

\begin{figure}[htbp]
\centering\includegraphics[width=12cm]{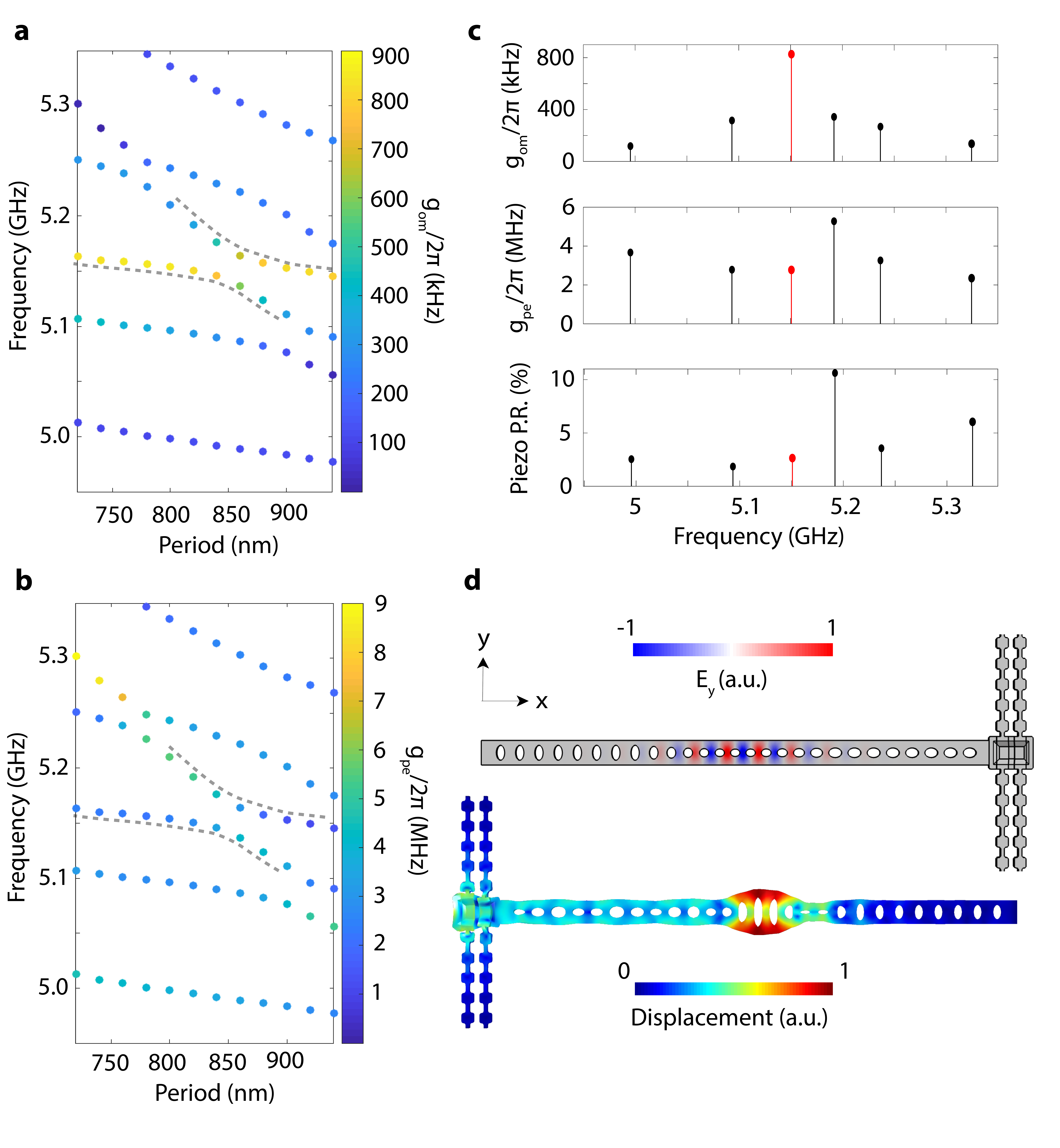}
\caption{a) Full transducer optomechanical coupling and mode structure vs. IDT period. Shown are all modes which have high piezoelectric and optomechanical coupling. Data points are colored according to $\gom$ value. Dashed lines follow the anticrossing of the maximum $\gom$ mode. b) Full transducer piezoelectric coupling and mode structure vs. IDT period. Shown are the same modes as in a), only colored according to $\gpe$. c) Mode structure at $p = 820$~nm, showing $\gpe$, $\gom$, and mechanical energy participation ratio in the piezoelectric region. Red shows mode with highest value of $\gom$. d) Mechanical mode profile of the mode highlighted in red in c).}
\label{fig:6}
\end{figure}

After independently designing the piezoacoustic and optomechanical cavities, we connect the two as shown in Fig.~\ref{fig:6}d, and simulate the resulting hybridized mode structure. To observe the hybridization of the piezoacoustic and optomechanical modes, we sweep the IDT period in the piezo region to tune the piezoacoustic mode through the multiple optomechanical resonances. The results are shown in Fig.~\ref{fig:6}a and b. We find that over a frequency window $>250$~MHz, there is a large number of mechanical modes with simultaneous high piezoelectric and optomechanical coupling rates. The phonon waveguide allows for strong enough mode hybridization that the piezoelectric coupling is distributed across a large number of modes. As shown in Fig.~\ref{fig:6}c, the mechanical energy participation in the piezo region $\zeta_m$ is in the range $1$-$10$\%. We find that across the entire hybridization window, at least one mode can be identified with $\gom/2\pi > 650$~kHz, $\gpe/2\pi > 1$~MHz, and $\zeta_m < 5$\%. In Fig.~\ref{fig:6}c, this mode is highlighted in red with $\gom/2\pi = 826$~kHz, $\gpe/2\pi = 2.8$~MHz, and $\zeta_m = 2.3$\%. We will use the values from this mode to quantify further calculations in this work.   
In practice, the frequencies and couplings of these mechanical modes are subject to change due to multiple sources of fabrication disorder. The multi-mode structure and relatively large hybridization ensure that the full device is robust to these shifts. While the exact frequencies and couplings may shift, Fig.~\ref{fig:6}a and \ref{fig:6}b illustrate that the qualitative nature of the mode structure remains unchanged for a large range of frequency shifts. Additionally, the modes are separated far enough in frequency that their parasitic effect on each other's transduction efficiencies is minimal.

We may use the simulated piezo participation ratio and radiation loss to estimate the mechanical decoherence rate $\kappa_m$ of our device. There are two dominant contributions to decoherence in our design. The first is acoustic radiation loss into the surrounding substrate. This can be simulated and is found to be in the range $\kappa_{\mathrm{rad}}/2\pi \sim 1-10$~kHz for all modes in Fig.~\ref{fig:6}, with $\kappa_{\mathrm{rad}}/2\pi = 2.3$~kHz for the mode highlighted in Fig.~\ref{fig:6}c. The second is coupling to two-level systems (TLS), which in both lithium niobate~\cite{LNAcousticLossTangent} and silicon~\cite{GregHighQAcoustics} has been found to be the dominant loss mechanism for GHz-frequency acoustic cavities at single phonon level powers and milliKelvin temperatures. For mechanical piezo participation ratio $\zeta_m$, the TLS decoherence rate can be estimated by $\kappa_{\mathrm{TLS}} = \zeta_m\kappa_{\mathrm{LN}} + (1 - \zeta_m)\kappa_{\mathrm{Si}}$. Using reported TLS-limited linewidths, $\kappa_{\mathrm{LN}}/2\pi \sim 300$~kHz in lithium niobate \cite{LNAcousticLossTangent} and $\kappa_{\mathrm{Si}}/2\pi \sim 4$~kHz in silicon \cite{GregHighQAcoustics}, and taking $\zeta_m$ = 2\%, we estimate a TLS decoherence rate of $\kappa_{\mathrm{TLS}}/2\pi \sim 10$~kHz. The total mechanical decoherence rate is then estimated to be in the range $\kappa_m/2\pi \sim 10-20$~kHz. 

\section{Efficiency and Added Noise}
\label{sec:5}

To analyze the efficiency and noise of our design, we consider a pulsed scheme for microwave-to-optical state transfer on a transmon qubit connected to the transducer~\cite{PainterTransduction}. The qubit is first tuned on resonance with the mechanical mode for a time $t = \pi/2\gpe$ to complete a microwave photon-phonon swap operation, and subsequently detuned far off-resonance. A red-detuned ($\omega_d - \omega_o = -\omega_m$) laser pulse is then used to upconvert this phonon into an optical photon. The intrinsic efficiency of such a pulsed scheme is simply given by $\eta_i = \etape\etaom$, where $\etape$ is the piezoelectric photon-phonon swap efficiency, $\etaom$ is the optomechanical phonon-photon conversion efficiency.

The photon-phonon swap efficiency, $\etape$, can be calculated from a master equation simulation of the qubit-mechanics system with the Hamiltonian $\hat{H}_{\text{pe}}$ described in Sec.~\ref{sec:1}. Using $\gpe/2\pi = 2.8$~MHz, an estimated $\kappa_q/2\pi = 50$~kHz from Sec.~\ref{sec:2}, and an estimated mechanics decoherence rate $\kappa_m/2\pi = 20$kHz from Sec.~\ref{sec:4}, we find $\etape$ = 95\%. The optomechanical readout step determines both $\etaom$ and the dominant noise contribution to the transducer, which arises from optical absorption heating of the mechanical mode.
For a laser pulse duration $\tau$, $\etaom$ is given by~\cite{PainterTransduction}, 

\begin{equation}
    \etaom (\tau) =  \frac{\gammaom}{\gammaom + \kappa_m} (1 - e^{-(\gammaom + \kappa_m)\tau}), 
    \label{eq:4}
\end{equation}

\noindent where $\gammaom = 4\gom^2 n_o/\kappa_o$ is the optomechanical scattering rate, and $n_o$ is the number of intracavity optical photons corresponding to peak power of the optical pulse. In principle, this efficiency may be unity in the limit $\tau \gg 1/(\gammaom + \kappa_m)$ and $\gammaom \gg \kappa_m$. However, optically-induced heating of the mechanical mode severely limits $\tau$ in order to maintain $<1$ added noise photon. This leads to a fundamental tradeoff between efficiency and added noise resulting from OMC heating dynamics at milliKelvin temperatures. Maximizing efficiency for a given level of added noise requires careful choice of pulse duration $\tau$ and optical power $n_o$.

The added noise phonons $n_m(\tau)$ during optical readout are thought to originate from optical excitation of material defect states which undergo phonon-assisted relaxation, thereby heating the mechanical mode of interest \cite{SMeenehan, GregHighQAcoustics}. The timescale $\tau_h$ for $n_m$ to exceed 1 noise phonon depends strongly on $n_o$ and is found to vary greatly in different devices. Experiments with low-loss ($\kappa_m \lesssim 10$~kHz) pure silicon OMCs report $\tau_h \sim 1~\mu$s \cite{SMeenehan, GregHighQAcoustics, Groblacher_HBT}, whereas those with silicon OMCs integrated in a piezo-optomechanical transducer with $\kappa_m = 1$~MHz report much shorter $\tau_h \sim 100$~ns~\cite{PainterTransduction}. This suggests the presence of additional sources of optically induced heating and mechanical damping in piezo-optomechanical transducers that are potentially correlated. Possible sources are optical absorption by the IDT electrodes, TLSs in the piezo region, and surface defects in the OMC region from additional steps in the transducer fabrication process. While the dynamics of optically induced heating in piezo-optomechanical devices is a subject of future studies, it is clear that a transducer design aimed at improving optomechanical readout efficiency and noise should make the acoustic mode involved in the transduction process as silicon-like as possible, since this is the lowest loss material in the transducer.

In the design presented above, we minimize the dimensions of the piezoacoustic cavity so that most of the energy in the mechanical mode lives in the OMC region. The estimated mechanical damping rates based on participation ratios of various regions and calculated optomechanical coupling rates are comparable to those realized in pure silicon OMCs. Therefore, we may approximate the heating dynamics of our design as similar to that reported in previous silicon OMC work \cite{SMeenehan, GregHighQAcoustics}. Using this heating model we estimate $\sim$0.5 added noise photons for a pulse with $n_o = 45$ and $\tau = 500$~ns. Using the previously estimated values $\kappa_m/2\pi = 20$~kHz, $\kappa_o/2\pi = 800$~MHz, and $\gom/2\pi = 826$~kHz (yielding $\gammaom/2\pi = 153$~kHz), we estimate a pulse with $n_o = 45$ and $\tau = 500$~ns can achieve $\etaom = 37$\%. Combined with $\etape = 95\%$, the estimated intrinsic efficiency is $\eta_i \sim 35$\%. There are additional noise sources which we have not considered here such as photodetector dark counts and residual photons from the optical pump pulse. However, given the measured photon count rates for these noise sources in previous milliKelvin optomechanics experiments \cite{SMeenehan}, these noise contributions are negligible compared to the one from optical absorption heating.

The total efficiency of the transducer includes efficiencies in the extraction and collection of optical photons produced in the transduction process, and is given by $\eta = \eta_i \eta_k \etaext$. Here, $\eta_k = (\kappa_{o, e}/\kappa_o)$ determines the fraction of optical photons emitted into the coupling waveguide, and $\etaext$ is the external photon collection efficiency. For $\kappa_{o, e} \approx \kappa_{o, i}$ (as is the case for typical 1D silicon OMC designs \cite{SMeenehan}) we have $\eta_k \approx 50\%$. In typical optomechanics experiments, $\etaext$ is mainly determined by the fiber-to-device coupling efficiency, insertion loss of the optical pump filtering setup, and quantum efficiency of single photon detectors. Based on values from recent experiments \cite{PainterTransduction, ASN_HeraldedPhotons, QPhoX_Transduction}, these factors can be estimated to be 60\%, 20\%, and 90\%, respectively, leading to $\etaext \sim 10\%$. The product of all three efficiency estimates above yields a total transducer conversion efficiency from microwave to optical photons of $\eta \sim 2$\%. 

Finally, we consider the expected repetition rate for the transduction sequence in the pulsed scheme described above. In previous work with similar piezo-optomechanical transducers~\cite{PainterTransduction}, this was limited to $100$~Hz by the $\sim$10-ms timescale for quasiparticle relaxation in the aluminum transmon coupled to the transducer.
We expect that using qubits and resonators fabricated in superconductors such as niobium and niobium nitride, with quasiparticle relaxation timescale in the $\sim$ns range~\cite{QPLifetime_Nb, QPLifetime_NbN}, will allow for repetition rates in the $10$~kHz range limited by the decay rate $\kappa_{\mathrm{rad}}$ of added noise phonons in the acoustic mode after the optical pulse. At this repetition rate and estimated total efficiency $\eta \sim 2$\%, we expect a single photon count rate of order $200$~Hz, and a photon coincidence rate of order $4$~Hz. These rates are a key figure of merit for heralded remote entanglement generation schemes \cite{FiguresOfMerit, DLCZ}, indicating that these sorts of experiments and applications are practically realizable with our proposed new transducer design. We note that the above estimates are calculated for microwave-to-optical frequency conversion and do not carry over to operation in the reverse direction due to the nature of parasitic heating in our device. However, this does not impact the utility of the device for remote entanglement of microwave qubits since we envision its use as a probabilistic microwave-to-optical frequency converter connected to a qubit. The resulting quantum correlations between the qubit and transduced optical photons can serve as a key resource to implement well-known probabilistic schemes for long-distance quantum communication based on optical photon detection \cite{DLCZ, FiguresOfMerit}. 

\section{Conclusion}
\label{sec:6}

We have presented an optimized design for an ultra-low mode volume piezo-optomechanical transducer. The choice of lithium niobate on silicon on insulator as the material platform and our design strategy, allow for transducer performance metrics in a parameter regime favorable for quantum network applications. Crucially, while these performance metrics are only valid for one-way conversion, this should not impact its effectiveness in remote qubit entanglement protocols. 

The efficiency, noise, and repetition rate of the above transducer design are expected to be limited by optomechanical heating rates. Future improvements can be made by employing 2D optomechanical crystal cavities~\cite{2DOMC}, which through better thermal conductance to the substrate, mitigate the impact of heating of the mechanical mode due to optical absorption. Further, the transmon qubit connected to the transducer can be replaced with a resonator made in  disordered superconductors such as NbN or NbTiN~\cite{NbNResonators, NbNResonators_CPW}, which provide high kinetic inductance and fast quasiparticle relaxation timescales. Such a transducer device could be connected to an off-chip qubit in a modular approach~\cite{JILA_Nondestructive}, which alleviates performance restrictions due to absorption of the transducer optical pump by superconducting qubits. With an order of magnitude higher impedance, kinetic inductance resonators can allow for further miniaturization of the piezo volume and the possibility of novel transducers with ultra-thin lithium niobate of thickness in the tens of nanometers range.\\ 


\begin{backmatter}


\bmsection{Acknowledgments}
 This work was supported by the ARO/LPS Cross Quantum Technology Systems program (grant
W911NF-18-1-0103), the U.S. Department of Energy Office of Science National Quantum Information
Science Research Centers (Q-NEXT, award DE-AC02-06CH11357), the Institute for Quantum Information
and Matter, an NSF Physics Frontiers Center (grant PHY-1125565) with support of the Gordon and Betty
Moore Foundation, the Kavli Nanoscience Institute at Caltech, and the AWS Center for Quantum Computing.
S.M. acknowledges support from the IQIM Postdoctoral Fellowship. The authors thank A. Sipahigil, M. Mirhosseini, and M.Kalaee for early contributions to this work, and S. Sonar and U. Hatipoglu for helpful discussions. 


\bmsection{Disclosures}
The authors declare no conflicts of interest.






\bmsection{Data Availability Statement}
Correspondence and requests for materials should be sent to OP (opainter@caltech.edu).

\end{backmatter}

\bibliography{sample}

\end{document}